\documentclass[11pt]{article}
\usepackage{graphicx}
\usepackage{color}
\usepackage{tabularx}
\usepackage{url}
\usepackage{comment}
\usepackage{caption}
\usepackage{multirow}

\usepackage[final]{acl}

\usepackage{times}
\usepackage{latexsym}

\usepackage[T1]{fontenc}

\usepackage[utf8]{inputenc}

\usepackage{microtype}

\usepackage{inconsolata}

\title{User Review Writing via Interview with Dialogue Systems}

\author{Yoshiki Tanaka \and Michimasa Inaba \\
  The University of Electro-Communications\\
\texttt{\{y-tanaka, m-inaba\}@uec.ac.jp}\\}

\begin{document}
\maketitle

\begin{abstract}
  User reviews on e-commerce and review sites are crucial for making purchase decisions, although creating detailed reviews is time-consuming and labor-intensive. In this study, we propose a novel use of dialogue systems to facilitate user review creation by generating reviews from information gathered during interview dialogues with users. To validate our approach, we implemented our system using GPT-4 and conducted comparative experiments from the perspectives of system users and review readers. The results indicate that participants who used our system rated their interactions positively. Additionally, reviews generated by our system required less editing to achieve user satisfaction compared to those by the baseline. We also evaluated the reviews from the readers' perspective and found that our system-generated reviews are more helpful than those written by humans. Despite challenges with the fluency of the generated reviews, our method offers a promising new approach to review writing.
\end{abstract}

\section{Introduction}
User reviews posted on e-commerce and review sites benefit both buyers and sellers. For users considering a purchase, these reviews provide product details and insights, thereby supporting the decision-making process of purchase. Additionally, they act as feedback for sellers, enabling product quality improvement. Reviews that contain specific product details or are based on personal user experiences are especially valuable. However, writing these insightful reviews can be time-consuming and labor-intensive, as the task requires careful language and proofreading for potential buyers.\par

\begin{figure}[t!]
	\begin{center}
		\includegraphics[width=\linewidth]{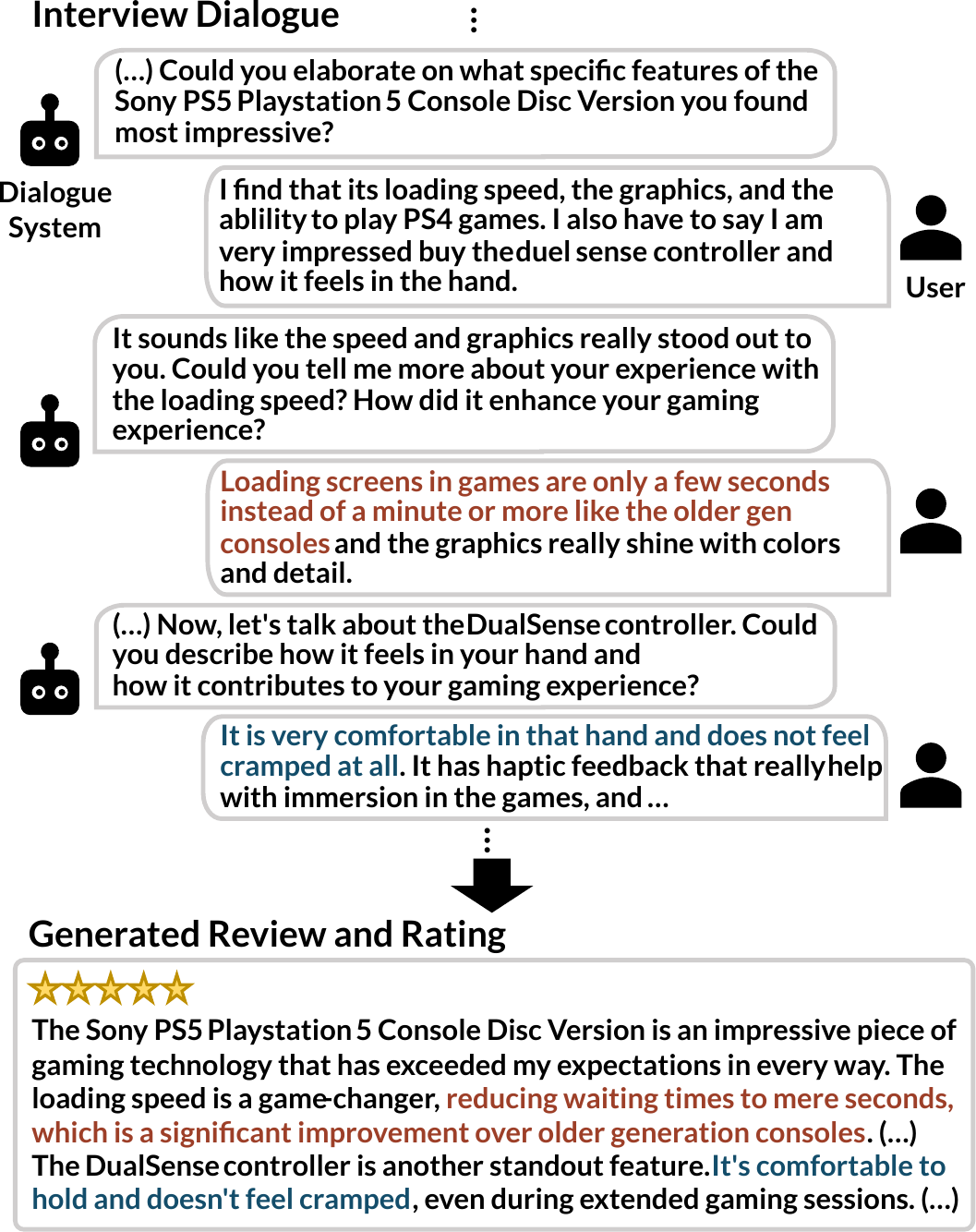}
		\caption{Example of review creation supported by the proposed method. In the dialogue and review, the contents of the same-colored text correspond to each other.}
		\label{fig:proposal}
	\end{center}
\end{figure}

To reduce users' burden of writing reviews, researchers (e.g., \citep{dong-etal-2017-learning-generate, li-tuzhilin-2019-towards}) have focused on automated review generation, thereby making the review creation process more efficient.\par

Despite these efforts, the ability of existing review generation systems to personalize for users and products remains limited. Previous studies have explored utilizing a wide range of information about products and users for generating reviews. However, available subjective information is limited. Specifically, without direct input from the user, incorporating the user's actual experiences with the product into the generated review is challenging. This constraint significantly limits the system's ability to personalize for the user. To overcome this problem, we focused on supporting the creation of reviews by directly eliciting information about products from users.\par

In this study, we propose the novel utilization of dialogue systems for creating user reviews. Figure \ref{fig:proposal} shows an example of the review creation process supported by the dialogue system according to our proposed method. First, the dialogue system acts as an interviewer, eliciting user opinions on products through interview dialogues.
Second, the review text generator generates review text based on the dialogue history. Finally, the rating predictor predicts a rating consistent with the generated review text. Our method allows users to easily create reviews by simply interacting with the system, thus reducing the effort involved in review creation.\par

To evaluate our method, we implemented a system incorporating our approach using GPT-4. Subsequently, we conducted experiments using our system, collecting data on dialogues between the system and users, the generated reviews, predicted ratings, and participants' feedback on our system. We discuss the effectiveness of our method after analyzing the collected data. In summary, our main contributions are as follows:

\begin{enumerate}
	\item As a novel application of dialogue systems, we propose a method for supporting user review creation. Furthermore, we developed a system incorporating our approach using GPT-4.
	\item We conducted a comprehensive survey from the perspectives of system users and review readers, showing that our method can provide high-quality and helpful reviews for both parties.
\end{enumerate}

\section{Related Work}

\subsection{Interview Dialogue System and Dataset}\label{sec:related-interview}
The interview dialogues are aimed at eliciting information from the interviewees. Prior research suggests that surveys conducted on chatbot platforms yield higher-quality responses than web survey platforms \citep{kim-2019-comparing}. This finding indicates that employing dialogue systems to collect user opinions and impressions is a promising approach.\par
Researchers have collected interview dialogue data on various topics, including radio \cite{majumder-etal-2020-interview}, news \cite{zhu-etal-2021-mediasum}, sports \cite{sun-etal-2022-sportsinterview}, and cooking \cite{okahisa-etal-2022-constructing}. The objectives of these collections vary from analyzing dialogue patterns \cite{majumder-etal-2020-interview, okahisa-etal-2022-constructing} to dialogue summarization \cite{zhu-etal-2021-mediasum}. Here, we utilize the interview dialogue system to support the creation of helpful reviews.

\subsection{Review Generation}
User reviews reflect user's opinions and requests regarding a product. These insights benefit buyers and sellers. Additionally, user reviews have a wide range of applications. Previous research has applied reviews to natural language processing tasks such as recommendations \cite{qiu-etal-2021-ubert}, opinion summarization \cite{brazinskas-etal-2020-unsupervised}, and task-oriented dialogue \cite{zhao-etal-2023-others}.\par
User reviews that include detailed information about the product and user experiences are useful. However, writing these reviews is a labor-intensive task for humans. To increase the efficiency of this process, researchers have proposed automated review generation models, enhancing their review generation capabilities by utilizing information such as ratings \cite{dong-etal-2017-learning-generate, sharma-etal-2018-cyclegen, li-etal-2019-generating, kim-etal-2020-retrieval}, images \cite{truong-etal-2019-multimodal, vu-etal-2020-multimodal}, past reviews written by the user \cite{li-tuzhilin-2019-towards}, and aspect-oriented features \cite{li-tuzhilin-2019-towards}. Unlike these studies, we focus on the collaborative writing of user reviews with the support of the dialogue system. \par

Some researchers have focused on supporting users in creating reviews, similar to our approach \cite{ni-mcauley-2018-personalized, bhat-etal-2023-hayes}. For example, Ni and McAuley proposed utilizing short phrases related to products that are provided by customers, such as review summaries and product titles, as auxiliary data for generating reviews \cite{ni-mcauley-2018-personalized}. In their system, the user provides information in a unidirectional manner. In contrast, we utilize an interview-specific dialogue system to collect information from the user through interactive interaction. The dialogue system can ask follow-up questions to obtain additional details regarding a product although this information may be ambiguous. This capability supports the creation of detailed reviews.\par

\subsection{Dialogue Summarization}
In our method, we proposed to convert conversational data (i.e., interview dialogue history) into non-conversational data (i.e., review texts). Therefore, our work is closely related to dialogue summarization research.
To build an effective model for dialogue summarization, researchers have proposed diverse approaches to learning methods \cite{zou-etal-2021-dams, li-etal-2023-dionysus, zhong-etal-2022-dialoglm, zhang-etal-2021-sutat}.
Additionally, researchers have built dialogue summarization datasets that can be used for training models; these datasets cover daily life conversations \cite{gliwa-etal-2019-samsum, chen-etal-2021-dialogsum}, meetings \cite{carletta-etal-2006-ami, zhong-etal-2021-qmsum}, TV series \cite{chen-etal-2022-summscreen}, and media dialogue \cite{zhu-etal-2021-mediasum}. While these studies aim to condense dialogue histories into brief texts, our approach takes a different direction. We focus on extracting useful product information for readers from interview dialogues and organizing it into a non-conversational data format, rather than compressing it into shorter text.

\section{Methodology}
\begin{figure}[t!]
	\begin{center}
		\includegraphics[width=\linewidth]{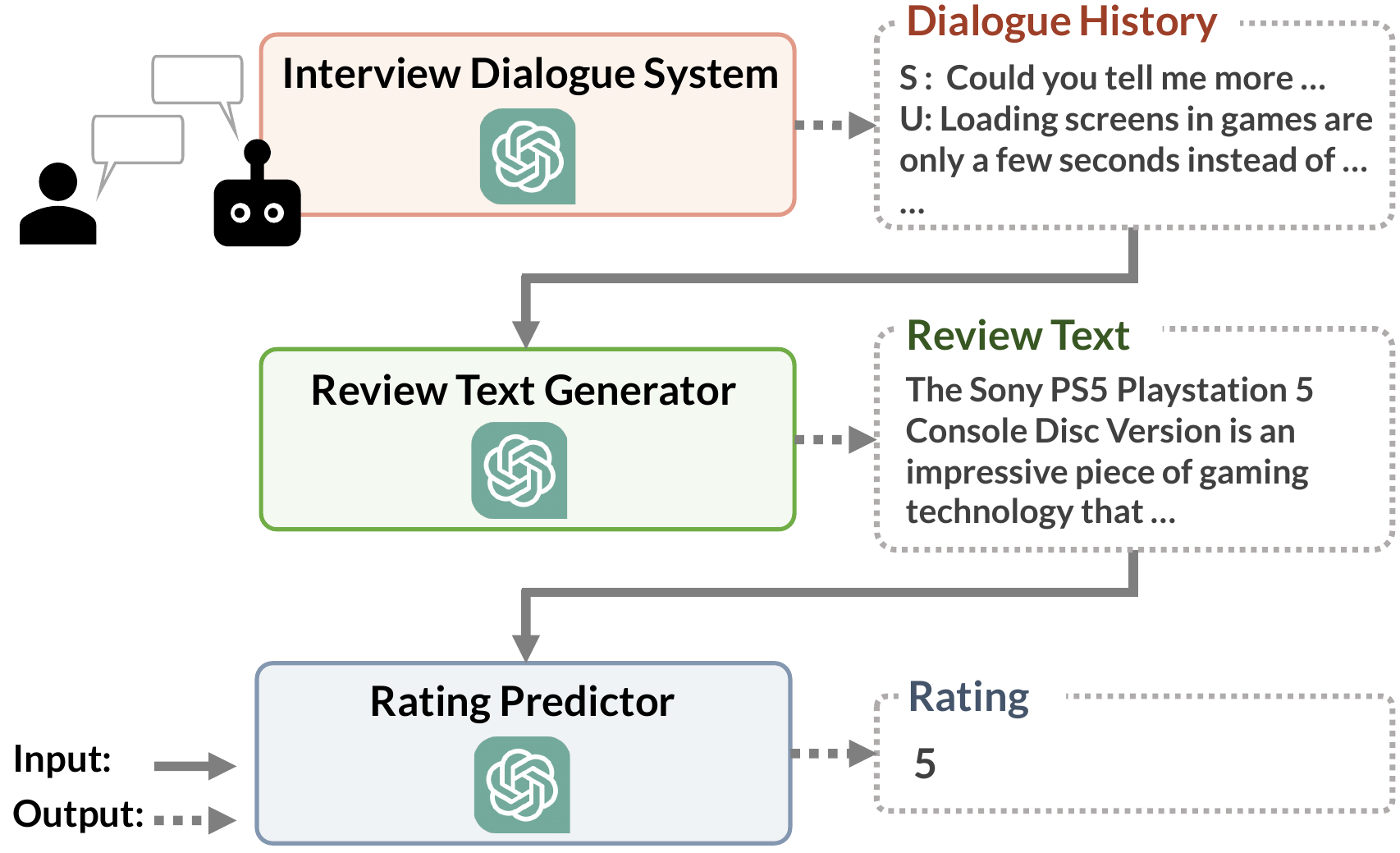}
		\caption{Overview of our system. First, the interview dialogue system interviews the user to elicit their impressions and requests about the product they used. Next, the review text generator uses the dialogue history as input to generate a review text. Finally, the rating predictor predicts a rating consistent with the sentiment of the generated review text.}
		\label{fig:system}
	\end{center}
\end{figure}

To create useful reviews, reviewers must provide detailed product information. Interview dialogue systems are employed to effectively elicit this information. To enhance readability, we propose organizing the dialogue history into a non-dialogue format.
Our method comprises three processes: interview dialogue, review text generation, and rating prediction. In this paper, the systems that perform these processes are referred to as the ``interview dialogue system,'' the ``review text generator,'' and the ``rating predictor,'' respectively. Our system utilizes these components in sequence to generate reviews as the output. An overview of our system is shown in Figure \ref{fig:system}. We use the gpt-4-0613 model to implement our system.\par

\subsection{Interview Dialogue System}
To assist potential buyers in making purchase decisions, guiding users to create helpful reviews is crucial. In our approach, therefore, our system should be designed to effectively collect information from the user. To achieve this, we propose utilizing an interview dialogue system. For the interview dialogue system, it is desirable to elicit both the pros and cons of a product in a balanced and detailed manner. Specifically, the system should be capable of asking follow-up questions about the content mentioned by the user or changing the topic to inquire about different aspects of the product.\par

We designed a prompt that incorporated instructions for the system to perform these behaviors. Moreover, aiming to both collect sufficient information for creating reviews and ensure users don't become bored, we added constraints regarding the number of turns to the prompt. In our experiments, we adopted instructions to ask at least 8 questions and conclude the interview within 15 turns. Additionally, to ensure the interview does not continue indefinitely, we externally implemented a setting in the interview dialogue system to end the dialogue after 15 turns. The prompt template for the interview dialogue is shown in Appendix \ref{sec:appendix-a-interview}.\par

\subsection{Review Text Generator}
Although the dialogue history between interview dialogue systems and users offers useful and detailed product information, it often contains redundancies. Consequently, it is not appropriate to post it directly as a user review. Therefore, we propose transforming the dialogue history into a format suitable for reviews. Our review text generator aims to capture the essence of the interview dialogue history while generating review texts from the perspective of the user. To generate reviews that align with the user's feedback, the system must faithfully reflect the content of the dialogue history in the review text. Our prompts include instructions to concisely summarize important information mentioned during the interview and generate the main body of the user review. The prompt template for the review generation is shown in Appendix \ref{sec:appendix-a-reviewgen}.

\subsection{Rating Predictor}
In e-commerce and review platforms, customer ratings are aggregated into a single score, providing other users with an initial impression of the product. For an aggregated score to be reliable, reviewers must assign ratings that accurately reflect the content of their review text. While the ratings impact the reputation widely, considering the potential for human error in assigning ratings, automating the task might be an effective solution. Our rating predictor automatically outputs a rating consistent with the sentiment of the input review text, ranging from 1 to 5 as an integer. Ratings consistent with the content of the review texts could reduce exaggerated scoring caused by user subjectivity. As a result, this can improve the reliability of the ratings.\par
We utilized GPT-4 to implement a rating predictor. To enhance predictive performance, we designed prompts that apply chain-of-thought prompting \cite{wei-etal-2022-cot, wang-etal-2022-selfconsistency, kojima-etal-2022-0shotcot}, that feeds large language models not only examples of question-and-answer pairs but also examples of the thought processes leading to those answers. In this study, we collected five sets of product titles, review texts, and ratings from Amazon.com to create output exemplars, each corresponding to ratings from 1 to 5. Subsequently, for each set, we wrote descriptions of the reasoning paths leading to the prediction of the rating from the product title and review text. We used these as few-shot exemplars within the context. Similarly, for target reviews, GPT-4 is encouraged to generate a reasoning path and an answer.

\section{Experiments}
We aim to facilitate the review-writing process for reviewers and provide helpful reviews to readers. To investigate the practicality of our method, we conducted evaluations from the perspectives of system users (Section \ref{sec:exp-collection}) and review readers (Section \ref{sec:exp-thirdparty}).

\subsection{Participant Evaluation}\label{sec:exp-collection}
To evaluate our method, we collected feedback through interviews, generated reviews, and questionnaires. Data collection was conducted through Amazon Mechanical Turk (MTurk)\footnote{\url{https://www.mturk.com}}.

\begin{figure}[t!]
  \begin{center}
      \includegraphics[width=1.0\linewidth]{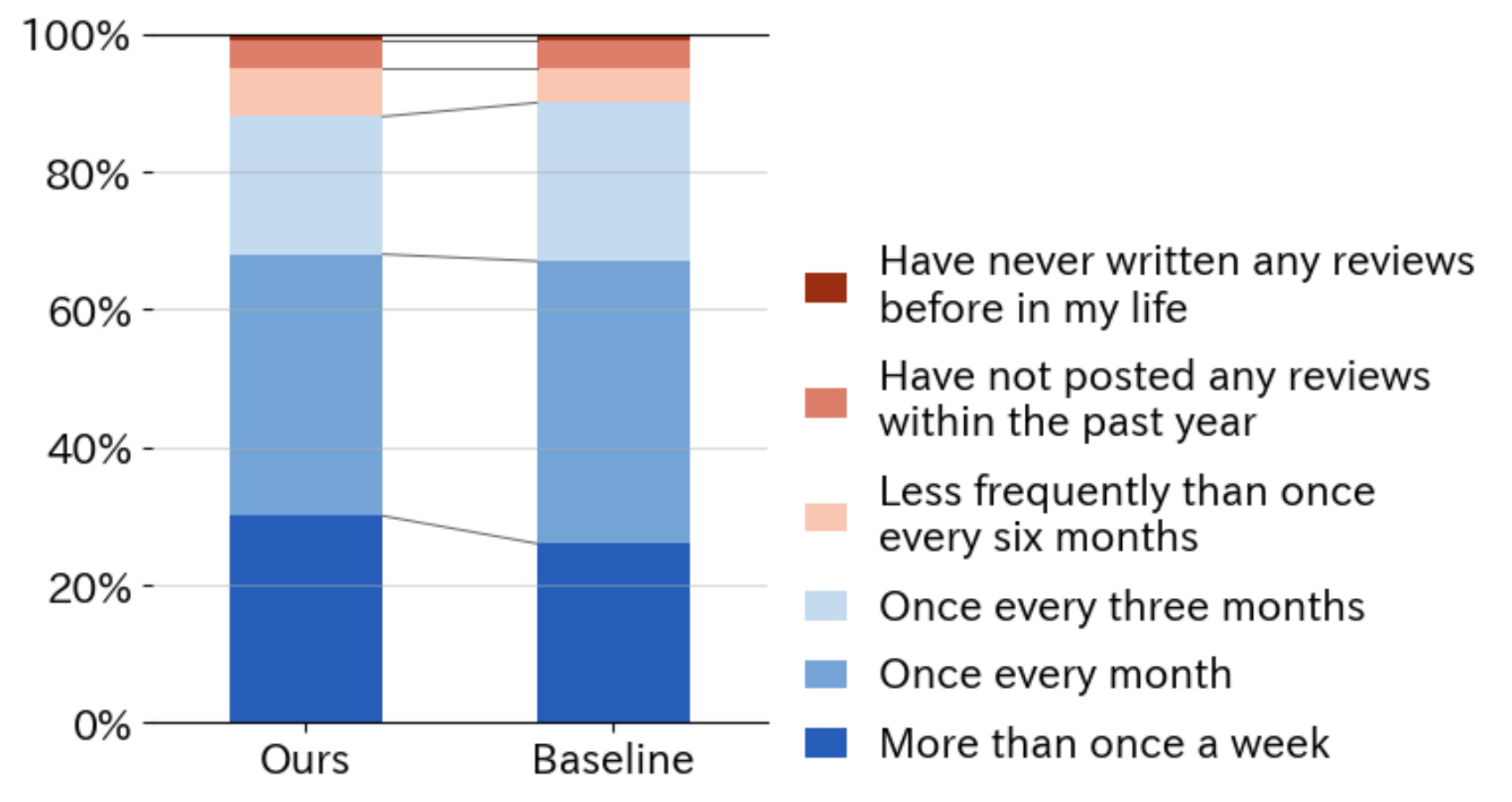}
      \caption{Participant responses to ``\textit{In the past year, how often have you posted reviews?}''}
      \label{fig:stat_post_history}
  \end{center}
\end{figure}

\subsubsection{Experimental Setup}
We tuned the temperature for each system. For the interview dialogue system, the temperature was set to 0.2. The review text generator and rating predictor generate outputs that are faithful to the input. Therefore, we set the temperature to 0 for these systems to suppress the diversity of the generated text.

\begin{table*}[t!]
	\centering
	\caption{Likert Items in Post-interview Survey}
  \scalebox{1.0}{
    \small
    \renewcommand{\arraystretch}{1.1}
    \begin{tabularx}{\linewidth}{l|l|X}\hline
      Dimension & Labels in Figure \ref{fig:likert_en} & Statements\\\hline
      \multirow{3}{*}{Interview} & Enjoyable & \textit{How fun was your interaction with the chatbot interviewer?}\\
      & Skillful & \textit{The interviewer skillfully elicited your impressions or opinions.}\\
      & In-depth& \textit{The chatbot interviewer attempted to elicit your impressions or opinions in depth.}\\\hline
      \multirow{2}{*}{Review} & Faithful & \textit{The system-generated review faithfully reflects what you said during your interviews.}\\
      & Concise& \textit{The system-generated review offers a concise summary of the points you mentioned during the interview.}\\\hline
      \multirow{3}{*}{System} & Quality & \textit{Please rate the overall quality of the system.}\\
      & Burdened(I) & \textit{I felt burdened to have an interview chat about the product.}\\
      & Burdened(R) & \textit{Writing a review with the support of the system is more burdensome than writing a review yourself.}\\
      \hline
    \end{tabularx}
  }
  \label{tb:post-survey}
\end{table*}

\subsubsection{Baseline System}
To demonstrate the effectiveness of using interview dialogue systems that adapt questions based on the context, we constructed a baseline system. The baseline system replaces the interview dialogue system with one that asks manually created questions in a fixed order. To construct the baseline system, we manually created nine questions on topics such as the reason for purchasing the product and the evaluation of the product in comparison with other products. All questions asked by the baseline system are listed in Appendix \ref{sec:appendix-b}. We collected data using this system in the same manner as with our proposed system.\par

\subsubsection{Evaluation Procedure}
Initially, participants conducted an interview dialogue with our interview dialogue system. After the interview, they were presented with the generated reviews and ratings. Participants then completed a post-interview survey comprising multiple-choice and open-ended questions. For each setting, we recruited 100 participants located in AU, CA, NZ, GB, or the US and had a 95\% approval rate with at least 500 previously approved HITs.\par

\begin{figure}[t!]
  \begin{center}
      \includegraphics[width=1.0\linewidth]{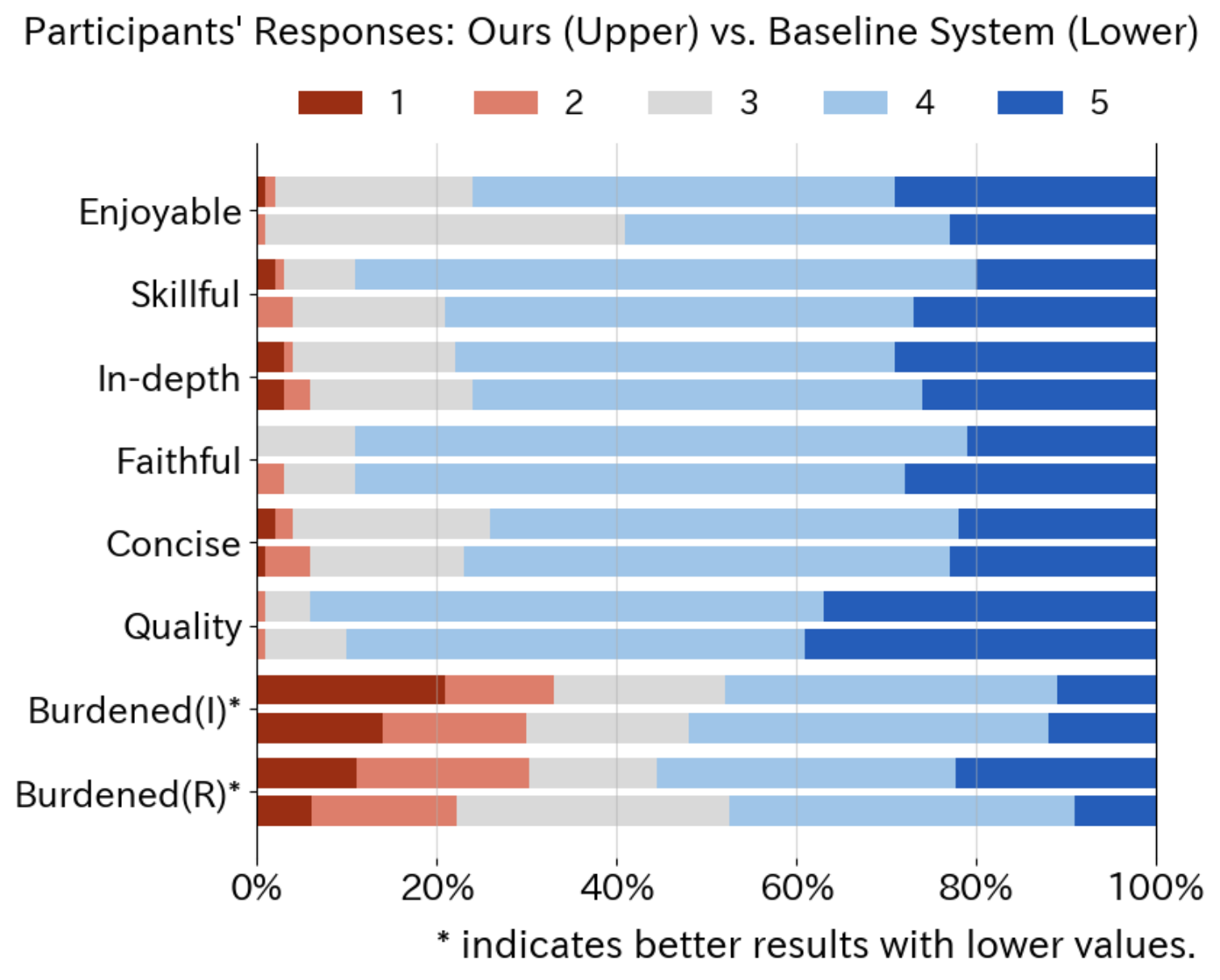}
      \caption{Participant responses to questions on a Likert scale from 1 (Strongly disagree) to 5 (Strongly agree) in a post-interview survey. For each question, the upper bar shows the results from our system and the lower bar shows the baseline results.}
      \label{fig:likert_en}
  \end{center}
\end{figure}

\subsubsection{Post-Interview Survey}\label{sec:post-interview}
After the interview, participants responded to a post-interview survey. Several questions in this survey were answered using a 5-point Likert scale. These questions are related to the interview dialogue, the generated reviews, and the overall system (See Table \ref{tb:post-survey}).\par
We also asked participants how frequently they post reviews to compare with their usual review-writing experiences. As shown in Figure \ref{fig:stat_post_history}, 95\% of the participants posted at least one user review in the past year. Additionally, participants were asked to rate the product they selected by responding to the question: "\textit{If you were to rate this product again, what rating would you give it?}" and provided a rating from 1 to 5.\par

\subsubsection{Participant Feedback Analysis}
Figure \ref{fig:likert_en} illustrates the distribution of responses to eight questions\footnote{For Burdened(R), we excluded responses from participants who selected the ``\textit{Have never written any reviews before in my life}'' option to the question in Figure \ref{fig:stat_post_history}.}. Regarding the dimensions of the interview and review, most participants evaluated two components positively: our interview dialogue system and our review text generator. Participants showed a similar positive trend across two settings for the four items: In-depth, Faithful, Concise, and Quality. Notably, for Quality, 90\% or more of the participants rated the overall quality of the systems positively.\par

\begin{figure}[t!]
  \begin{center}
      \includegraphics[width=1.0\linewidth]{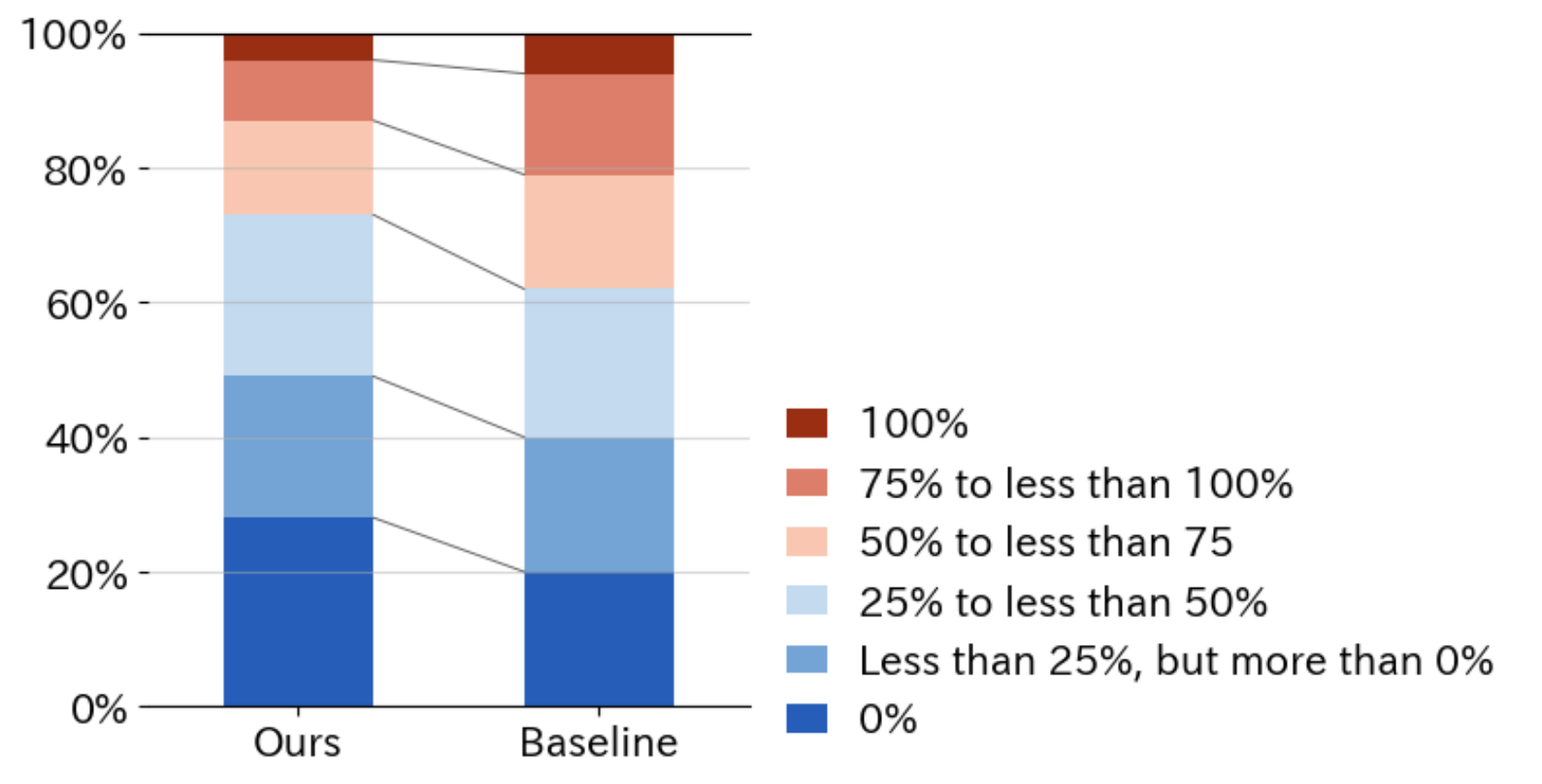}
      \caption{Participant responses to ``\textit{If you had to edit and post a system-generated review to your satisfaction, how much of it would you need to rewrite?}''}
      \label{fig:rewrite}
  \end{center}
\end{figure}

Our system provided users with more enjoyable interviews and higher satisfaction regarding the generated reviews compared with the baseline system. As shown in Figure \ref{fig:likert_en}, when using our system based on GPT-4, more participants agreed that interacting with the system was fun. Moreover, the difference in the methods used to elicit information—our interview dialogue system and the baseline—impacts users' enjoyment, with statistically significant differences (Mann–Whitney U test, $p < 0.05$). Participants also responded to the multiple-choice question, ``\textit{If you had to edit and post a system-generated review to your satisfaction, how much of it would you need to rewrite?}''. Figure \ref{fig:rewrite} shows that different types of systems resulted in varied response distributions.
In particular, 38\% and 27\% of participants using the baseline system and our system, respectively, responded that they needed to rewrite more than 50\% of the review. These results indicate that our system can provide reviews with higher satisfaction than the baseline.\par

Our system imposed a greater burden on participants. Figure \ref{fig:likert_en} shows that a higher percentage of participants agreed that \textit{writing a review with the support of our system is more burdensome than writing alone}, compared to the baseline. We argue that the response time of the system is one of the reasons for this difference.
Our GPT-4-based system, which generates responses based on users' utterances, takes a longer time to generate responses than a baseline that asks predefined questions. Notably, several participants suggested that the response speed of our system should be improved. In response to the free-form question ``\textit{What is one enhancement that can be made to improve this system?}'', we received answers such as ``\textit{more fast replies}'' and ``\textit{need quick reply}.'' In our experiments, unlike the ChatGPT interface\footnote{\url{https://chatgpt.com/}}, we did not employ real-time response generation using streaming functionality. Adding this feature would be an effective modification to enhance our system's response speed, which is expected to significantly improve user experience.

\begin{table*}[t!]
\centering
\caption{Example of a dialogue between our system and the user (top, S: system, U: user) and generated review text (bottom). In this example, our rating predictor predicted a rating of 4.}
\scalebox{1.0}{
  \setlength{\tabcolsep}{2mm}
  \small
  \renewcommand{\arraystretch}{1.1}
  \begin{tabularx}{\linewidth}{lX}\hline
    S&(...) To start off, could you tell me about your overall satisfaction with this product?\\
    U&\textbf{i would say well satisfied} but with few minor issues.\\
    S&I see, you're generally satisfied but there are a few minor issues. Could you elaborate on what you particularly like about the Braun Series 9 9370cc Electric Shaver?\\
    U&\textbf{some times small hair from the beard gets stucks} in particular spot on the blade area in the machine.\\
    S&I understand that small hairs getting stuck in the blade area is one of the issues you've encountered. But before we delve into that, could you first share what you find satisfying or impressive about the product? What are its positive points in your opinion?\\
    U&the postivie sides is smooth shaving experience and battery capacity which is good\\
    S&It's great to hear that you appreciate the smooth shaving experience and the battery capacity. Now, let's go back to the issue you mentioned earlier about small hairs getting stuck. Could you tell me more about this problem? How often does it happen and how does it affect your shaving experience?\\
    U&i wouls say not often, ifaced this issue only once.\\
    S&I see, (...) That's good to know. Now, aside from the smooth shaving experience and the battery capacity, are there any other features of the Braun Series 9 9370cc Electric Shaver that you find beneficial or impressive?\\
    &...\\\hline
    \multicolumn{2}{X}{\textbf{I am generally satisfied} with the Braun Series 9 9370cc Rechargeable Wet \& Dry Men's Electric Shaver with Clean \& Charge Station. The product offers a smooth shaving experience, which I find impressive. The battery capacity is also commendable, lasting for a good amount of time. I appreciate the versatility of the product, as it can be used for both wet and dry shaving. The Clean \& Charge Station that comes with the shaver is quite efficient and useful. It effectively cleans and charges the shaver, making it more convenient compared to other products I've used before. \textbf{However, I did encounter a minor issue where small hairs from my beard got stuck} in a particular spot on the blade area. This happened only once and did not significantly affect my overall shaving experience. Considering all the features and my experience with the product, I believe it offers good value for its price. I would recommend the Braun Series 9 9370cc Electric Shaver to others, mainly for its battery performance and its variety of features, including its wet and dry usage.}\\\hline
    \label{tb:example}
  \end{tabularx}
}
\end{table*}

\subsubsection{Case Study}
Our interview dialogue system can generate follow-up questions that explore the content of users' ambiguous responses in depth.
Table \ref{tb:example} shows an example of the data collected, comprising the dialogue history regarding an electric shaver and the corresponding review text generated. During the interview, our system initially asked about the participant's overall satisfaction with the product, to which the participant replied, ``\textit{... well satisfied but with few minor issues.}''
Based on this response, our system posed follow-up questions to clarify the aspects that the participant was satisfied with and the issues they mentioned. As demonstrated in this example, our system can elicit deeper information about products from users compared with the baseline system.\par

Additionally, our interview dialogue system can change topics during an interview to collect information on various aspects of a product. For the interview dialogues shown in Table \ref{tb:example}, the first three turns focus on the participant's satisfaction with the product. Subsequently, the system changes the topic by saying ``\textit{Now, let's go back to the issue}'' and thus shifting focus to the issues that the user mentioned in the first turn. In the next turn, our system shifts the topic again to highlight the product's impressive features. By switching topics in this manner, our system can acquire information on various aspects of the product.\par
Our system can also generate review texts and ratings that faithfully reflect the content mentioned in the interview. As shown in the lower part of Table \ref{tb:example}, the generated review text closely aligns with user comments such as ``i would say well satisfied ...'' and ``some times small hair from the beard gets stucks.'' Moreover, our system accounts for issues mentioned by the user and predicts a rating of 4, one less than the maximum.

\subsection{Third-Party Evaluation}\label{sec:exp-thirdparty}
Helpful user reviews offer effective decision-making support for potential buyers and sellers. To assess the quality of the reviews generated by our system from the reader's perspective, we conducted multidimensional evaluations.

\subsubsection{Experimental Procedure}
This evaluation was conducted in a pairwise setting. In this experiment, annotators were presented with two reviews: one generated by our system and another for comparison. We compared our system-generated reviews with Amazon reviews to evaluate the quality difference from human-written reviews. Additionally, to assess the differences from the perspective of the readers, we compared our system-generated reviews with those generated by a baseline system, which were collected in Section \ref{sec:exp-collection}.\par
Annotators chose either the superior option or a tie for each of the eight dimensions, as shown in Table \ref{tb:ques-comparison}. We recruited the annotators through MTurk.

\subsubsection{Selection Criteria for Comparative Reviews}
We collected human-written reviews from the Amazon Review Dataset \cite{ni-etal-2019-justifying}. To facilitate comparison for the annotators, we initially removed any human-written reviews where the category or rating did not match those generated by our system. After this filtering, to avoid selecting poor-quality reviews, we retained reviews in the top 5\% of human-written reviews based on helpfulness votes, using these as comparison candidates. Finally, among these candidates, we selected the review of the product whose title achieved the highest ROUGE-L score \cite{lin-2004-rouge}.\par
We applied the same process when matching reviews generated by the baseline system; however, for some reviews collected using our system, there were no corresponding reviews in the baseline data that matched both category and rating. Therefore, we excluded these unmatched data. Consequently, while we conducted comparisons for all 100 reviews written by humans, only 96 baseline-generated reviews met the criteria.

\subsubsection{Results and Discussions}
The overall results are presented in Table \ref{tb:comparison}. The annotators prefer the reviews generated by our system to those written by humans or generated by the baseline system.
Notably, the reviews generated by our system are helpful, provide a balanced view of pros and cons, and offer comprehensive information. These findings indicate that our interview dialogue system is capable of eliciting a wide range of information about products from users through topic transitions.\par

The reviews generated by our system lack the fluency of human-written reviews. For instance, our review text generator tends to use the formal product title when referring to the product. Additionally, human-written reviews contain more individual experiences compared with those generated by our system. Despite these limitations, our system has high scalability, offering the potential for improvement. Speciﬁcally, our system's output could be enhanced by reﬁning the prompts to generate texts that are more human-like and elicit detailed usage experiences from users.\par

By replacing the baseline system, which uses fixed questions, with our interview dialogue system, we observe improvements across all metrics. Notably, our system can generate reviews that are rich in experience-based information, contain more detailed information, and cover a broader range of topics. This demonstrates that our system can elicit more detailed and extensive information from users through follow-up questions and topic transitions.

\begin{table*}[t!]
	\centering
	\caption{Questions in comparative evaluation}
  \scalebox{1.0}{
    \small
    \renewcommand{\arraystretch}{1.1}
    \begin{tabularx}{0.95\linewidth}{l|X}\hline
      Labels in Table \ref{tb:comparison} & Questions\\\hline
      Helpfulness & \textit{Which review would be more helpful for making a purchase decision?}\\
      Fluency & \textit{Which review exhibits a more fluent and human-like writing style?}\\
      Conciseness & \textit{Which review is more concise and to the point?}\\
      Experience & \textit{Which review provides more information based on the actual usage experience of the product?}\\
      Balance & \textit{Which review presents a more balanced view of the product's pros and cons?}\\
      Depth & \textit{Which review provides more in-depth information about any specific aspect of the product?}\\
      Coverage & \textit{Which review mentions a more comprehensive range of product aspects?}\\
      Overall & \textit{Which review is overall more preferable?}\\
      \hline
    \end{tabularx}
  }
  \label{tb:ques-comparison}
\end{table*}

\begin{table*}[t!]
	\centering
	\caption{Results of third-party evaluation. The values represent the percentage of votes each received.}
  \scalebox{1.0}{
    \small
    \renewcommand{\arraystretch}{1.1}
    \begin{tabularx}{0.95\textwidth}{l|*{8}{>{\centering\arraybackslash}X}}\hline
      Reviews & Helpfulness & Fluency & Conciseness & Experience & Balance & Depth & Coverage & Overall\\\hline
      Human&38.0&\textbf{47.0}&37.0&\textbf{57.0}&37.0&43.0&40.0&41.0\\
      Tie&6.0&15.0&6.0&9.0&15.0&10.0&5.0&7.0\\
      Ours&\textbf{56.0}&38.0&\textbf{57.0}&34.0&\textbf{48.0}&\textbf{47.0}&\textbf{55.0}&\textbf{52.0}\\\hline
      Baseline&38.5&28.1&45.8&21.9&34.4&35.4&35.4&37.5\\
      Tie&12.5&33.3&6.2&16.7&15.6&12.5&10.4&17.7\\
      Ours&\textbf{49.0}&\textbf{38.5}&\textbf{47.9}&\textbf{61.5}&\textbf{50.0}&\textbf{52.1}&\textbf{54.2}&\textbf{44.8}\\\hline
    \end{tabularx}
  }
  \label{tb:comparison}
\end{table*}

\begin{table}[t!]
	\centering
	\caption{Average absolute difference in ratings between Amazon customers and Turkers (top-left), between system-predicted ratings and Turkers' ratings for system-generated reviews (top-right), and between system-predicted ratings and participants' ratings (bottom-right, see Section \ref{sec:post-interview}).}
  \small
  \renewcommand{\arraystretch}{1.2}
  \begin{tabular}{l|*{2}{>{\centering\arraybackslash}p{1.2cm}}}
    \hline
    Annotator/Source&Human-written&System-generated\\\hline
    Turkers&0.59&0.12\\
    Participants in Section \ref{sec:exp-collection}&-&0.57\\
    \hline
  \end{tabular}
  \label{tb:rating-statistics}
\end{table}

\subsection{Discussion on Predicted Ratings}
To further explore the characteristics of the reviews and ratings generated by our system, we analyze them along two axes: the difference based on the source of the ratings (comparing ratings assigned by humans to those predicted by our system) and the difference based on the annotators (comparing the ratings given by system users to those assigned by third parties). To obtain ratings assigned by third parties, we newly recruited annotators from MTurk and asked them to assign ratings to both the human-written reviews (left column)\footnote{For the annotations, we used 100 human-written reviews selected in Section \ref{sec:exp-collection}.} and those generated by our system (right column). We also collected ratings assigned by participants from the experiments in Section \ref{sec:exp-collection}. Note that these participants, unlike the Turkers, had seen the ratings predicted by our system.\par
The results in the top row of Table \ref{tb:rating-statistics} demonstrate that the difference between the ratings predicted by our system and those assigned by third parties is remarkably smaller than the difference found in human-written reviews. This finding indicates that the sentiment of the reviews generated by our system is easily comprehensible to readers.\par
The ratings predicted by our system, as shown in the right column of Table \ref{tb:rating-statistics}, align more closely with those assigned by third-party annotators than with those of system users. This finding indicates that our system emphasizes objectivity over subjectivity in its ratings.\par
The aforementioned observations indicate that our system generates review texts that are easy for humans to understand and provide more objective ratings. This finding suggests that our interview dialogue system and review text generator can generate reviews that accurately capture reviewers' sentiments, thereby supporting informed purchasing decisions, while the rating predictor also provides highly objective and reliable ratings.

\section{Conclusion}
In this study, we present a novel method for utilizing dialogue systems to facilitate user review creation. Our approach involves three processes: interview dialogue, review text generation, and rating prediction. Although ensuring the fluency of the system-generated reviews remains a challenge, our method provides high-quality and helpful reviews for both reviewers and their readers.\par

Our method possesses high scalability. For instance, feeding product descriptions into our interview dialogue system could lead to deeper interview dialogues about more detailed information. However, our experiments have shown that even without such extensions, our system is capable of providing reviews that are more helpful than human-written ones.
Furthermore, adapting our dialogue system's strategies to user preferences during review writing could improve user experience. Further research can accomplish this objective by conducting a more detailed analysis of user preferences.


\bibliographystyle{acl_natbib}
\bibliography{tanaka,custom}

\begin{thebibliography}{31}
\expandafter\ifx\csname natexlab\endcsname\relax\def\natexlab#1{#1}\fi

\bibitem[{Bhat et~al.(2023)Bhat, Agashe, Oberoi, Mohile, Jangir, and Joshi}]{bhat-etal-2023-hayes}
Advait Bhat, Saaket Agashe, Parth Oberoi, Niharika Mohile, Ravi Jangir, and Anirudha Joshi. 2023.
\newblock \href {https://doi.org/10.1145/3581641.3584060} {Interacting with next-phrase suggestions: How suggestion systems aid and influence the cognitive processes of writing}.
\newblock In \emph{Proceedings of the 28th International Conference on Intelligent User Interfaces}, IUI '23, page 436^^e2^^80^^93452, New York, NY, USA. Association for Computing Machinery.

\bibitem[{Bra{\v{z}}inskas et~al.(2020)Bra{\v{z}}inskas, Lapata, and Titov}]{brazinskas-etal-2020-unsupervised}
Arthur Bra{\v{z}}inskas, Mirella Lapata, and Ivan Titov. 2020.
\newblock \href {https://doi.org/10.18653/v1/2020.acl-main.461} {Unsupervised opinion summarization as copycat-review generation}.
\newblock In \emph{Proceedings of the 58th Annual Meeting of the Association for Computational Linguistics}, pages 5151--5169, Online. Association for Computational Linguistics.

\bibitem[{Carletta et~al.(2006)Carletta, Ashby, Bourban, Flynn, Guillemot, Hain, Kadlec, Karaiskos, Kraaij, Kronenthal, Lathoud, Lincoln, Lisowska, McCowan, Post, Reidsma, and Wellner}]{carletta-etal-2006-ami}
Jean Carletta, Simone Ashby, Sebastien Bourban, Mike Flynn, Mael Guillemot, Thomas Hain, Jaroslav Kadlec, Vasilis Karaiskos, Wessel Kraaij, Melissa Kronenthal, Guillaume Lathoud, Mike Lincoln, Agnes Lisowska, Iain McCowan, Wilfried Post, Dennis Reidsma, and Pierre Wellner. 2006.
\newblock The ami meeting corpus: A pre-announcement.
\newblock In \emph{Machine Learning for Multimodal Interaction}, pages 28--39, Berlin, Heidelberg. Springer Berlin Heidelberg.

\bibitem[{Chen et~al.(2022)Chen, Chu, Wiseman, and Gimpel}]{chen-etal-2022-summscreen}
Mingda Chen, Zewei Chu, Sam Wiseman, and Kevin Gimpel. 2022.
\newblock \href {https://doi.org/10.18653/v1/2022.acl-long.589} {{S}umm{S}creen: A dataset for abstractive screenplay summarization}.
\newblock In \emph{Proceedings of the 60th Annual Meeting of the Association for Computational Linguistics (Volume 1: Long Papers)}, pages 8602--8615, Dublin, Ireland. Association for Computational Linguistics.

\bibitem[{Chen et~al.(2021)Chen, Liu, and Zhang}]{chen-etal-2021-dialogsum}
Yulong Chen, Yang Liu, and Yue Zhang. 2021.
\newblock \href {https://doi.org/10.18653/v1/2021.inlg-1.33} {{D}ialog{S}um challenge: Summarizing real-life scenario dialogues}.
\newblock In \emph{Proceedings of the 14th International Conference on Natural Language Generation}, pages 308--313, Aberdeen, Scotland, UK. Association for Computational Linguistics.

\bibitem[{Dong et~al.(2017)Dong, Huang, Wei, Lapata, Zhou, and Xu}]{dong-etal-2017-learning-generate}
Li~Dong, Shaohan Huang, Furu Wei, Mirella Lapata, Ming Zhou, and Ke~Xu. 2017.
\newblock \href {https://aclanthology.org/E17-1059} {Learning to generate product reviews from attributes}.
\newblock In \emph{Proceedings of the 15th Conference of the {E}uropean Chapter of the Association for Computational Linguistics: Volume 1, Long Papers}, pages 623--632, Valencia, Spain. Association for Computational Linguistics.

\bibitem[{Gliwa et~al.(2019)Gliwa, Mochol, Biesek, and Wawer}]{gliwa-etal-2019-samsum}
Bogdan Gliwa, Iwona Mochol, Maciej Biesek, and Aleksander Wawer. 2019.
\newblock \href {https://doi.org/10.18653/v1/D19-5409} {{SAMS}um corpus: A human-annotated dialogue dataset for abstractive summarization}.
\newblock In \emph{Proceedings of the 2nd Workshop on New Frontiers in Summarization}, pages 70--79, Hong Kong, China. Association for Computational Linguistics.

\bibitem[{Kim et~al.(2020)Kim, Choi, Amplayo, and Hwang}]{kim-etal-2020-retrieval}
Jihyeok Kim, Seungtaek Choi, Reinald~Kim Amplayo, and Seung-won Hwang. 2020.
\newblock \href {https://doi.org/10.18653/v1/2020.coling-main.207} {Retrieval-augmented controllable review generation}.
\newblock In \emph{Proceedings of the 28th International Conference on Computational Linguistics}, pages 2284--2295, Barcelona, Spain (Online). International Committee on Computational Linguistics.

\bibitem[{Kim et~al.(2019)Kim, Lee, and Gweon}]{kim-2019-comparing}
Soomin Kim, Joonhwan Lee, and Gahgene Gweon. 2019.
\newblock \href {https://doi.org/10.1145/3290605.3300316} {Comparing data from chatbot and web surveys: Effects of platform and conversational style on survey response quality}.
\newblock In \emph{Proceedings of the 2019 CHI Conference on Human Factors in Computing Systems}, CHI '19, page 1^^e2^^80^^9312, New York, NY, USA. Association for Computing Machinery.

\bibitem[{Kojima et~al.(2022)Kojima, Gu, Reid, Matsuo, and Iwasawa}]{kojima-etal-2022-0shotcot}
Takeshi Kojima, Shixiang~(Shane) Gu, Machel Reid, Yutaka Matsuo, and Yusuke Iwasawa. 2022.
\newblock \href {https://proceedings.neurips.cc/paper_files/paper/2022/file/8bb0d291acd4acf06ef112099c16f326-Paper-Conference.pdf} {Large language models are zero-shot reasoners}.
\newblock In \emph{Advances in Neural Information Processing Systems}, volume~35, pages 22199--22213. Curran Associates, Inc.

\bibitem[{Li et~al.(2019)Li, Zhao, Wen, and Song}]{li-etal-2019-generating}
Junyi Li, Wayne~Xin Zhao, Ji-Rong Wen, and Yang Song. 2019.
\newblock \href {https://doi.org/10.18653/v1/P19-1190} {Generating long and informative reviews with aspect-aware coarse-to-fine decoding}.
\newblock In \emph{Proceedings of the 57th Annual Meeting of the Association for Computational Linguistics}, pages 1969--1979, Florence, Italy. Association for Computational Linguistics.

\bibitem[{Li and Tuzhilin(2019)}]{li-tuzhilin-2019-towards}
Pan Li and Alexander Tuzhilin. 2019.
\newblock \href {https://doi.org/10.18653/v1/D19-1319} {Towards controllable and personalized review generation}.
\newblock In \emph{Proceedings of the 2019 Conference on Empirical Methods in Natural Language Processing and the 9th International Joint Conference on Natural Language Processing (EMNLP-IJCNLP)}, pages 3237--3245, Hong Kong, China. Association for Computational Linguistics.

\bibitem[{Li et~al.(2023)Li, Peng, He, Galley, Yu, and Gao}]{li-etal-2023-dionysus}
Yu~Li, Baolin Peng, Pengcheng He, Michel Galley, Zhou Yu, and Jianfeng Gao. 2023.
\newblock \href {https://doi.org/10.18653/v1/2023.acl-long.76} {{DIONYSUS}: A pre-trained model for low-resource dialogue summarization}.
\newblock In \emph{Proceedings of the 61st Annual Meeting of the Association for Computational Linguistics (Volume 1: Long Papers)}, pages 1368--1386, Toronto, Canada. Association for Computational Linguistics.

\bibitem[{Lin(2004)}]{lin-2004-rouge}
Chin-Yew Lin. 2004.
\newblock \href {https://aclanthology.org/W04-1013} {{ROUGE}: A package for automatic evaluation of summaries}.
\newblock In \emph{Text Summarization Branches Out}, pages 74--81, Barcelona, Spain. Association for Computational Linguistics.

\bibitem[{Majumder et~al.(2020)Majumder, Li, Ni, and McAuley}]{majumder-etal-2020-interview}
Bodhisattwa~Prasad Majumder, Shuyang Li, Jianmo Ni, and Julian McAuley. 2020.
\newblock \href {https://doi.org/10.18653/v1/2020.emnlp-main.653} {Interview: Large-scale modeling of media dialog with discourse patterns and knowledge grounding}.
\newblock In \emph{Proceedings of the 2020 Conference on Empirical Methods in Natural Language Processing (EMNLP)}, pages 8129--8141, Online. Association for Computational Linguistics.

\bibitem[{Ni et~al.(2019)Ni, Li, and McAuley}]{ni-etal-2019-justifying}
Jianmo Ni, Jiacheng Li, and Julian McAuley. 2019.
\newblock \href {https://doi.org/10.18653/v1/D19-1018} {Justifying recommendations using distantly-labeled reviews and fine-grained aspects}.
\newblock In \emph{Proceedings of the 2019 Conference on Empirical Methods in Natural Language Processing and the 9th International Joint Conference on Natural Language Processing (EMNLP-IJCNLP)}, pages 188--197, Hong Kong, China. Association for Computational Linguistics.

\bibitem[{Ni and McAuley(2018)}]{ni-mcauley-2018-personalized}
Jianmo Ni and Julian McAuley. 2018.
\newblock \href {https://doi.org/10.18653/v1/P18-2112} {Personalized review generation by expanding phrases and attending on aspect-aware representations}.
\newblock In \emph{Proceedings of the 56th Annual Meeting of the Association for Computational Linguistics (Volume 2: Short Papers)}, pages 706--711, Melbourne, Australia. Association for Computational Linguistics.

\bibitem[{Okahisa et~al.(2022)Okahisa, Tanaka, Kodama, Huang, and Kurohashi}]{okahisa-etal-2022-constructing}
Taro Okahisa, Ribeka Tanaka, Takashi Kodama, Yin~Jou Huang, and Sadao Kurohashi. 2022.
\newblock \href {https://aclanthology.org/2022.lrec-1.335} {Constructing a culinary interview dialogue corpus with video conferencing tool}.
\newblock In \emph{Proceedings of the Thirteenth Language Resources and Evaluation Conference}, pages 3131--3139, Marseille, France. European Language Resources Association.

\bibitem[{Qiu et~al.(2021)Qiu, Wu, Gao, and Fan}]{qiu-etal-2021-ubert}
Zhaopeng Qiu, Xian Wu, Jingyue Gao, and Wei Fan. 2021.
\newblock \href {https://doi.org/10.1609/aaai.v35i5.16557} {U-bert: Pre-training user representations for improved recommendation}.
\newblock volume~35, pages 4320--4327.

\bibitem[{Sharma et~al.(2018)Sharma, Sharma, Bishnu, and Patel}]{sharma-etal-2018-cyclegen}
Vasu Sharma, Harsh Sharma, Ankita Bishnu, and Labhesh Patel. 2018.
\newblock \href {https://doi.org/10.18653/v1/W18-6552} {{C}yclegen: Cyclic consistency based product review generator from attributes}.
\newblock In \emph{Proceedings of the 11th International Conference on Natural Language Generation}, pages 426--430, Tilburg University, The Netherlands. Association for Computational Linguistics.

\bibitem[{Sun et~al.(2022)Sun, Cao, and Yang}]{sun-etal-2022-sportsinterview}
Hanfei Sun, Ziyuan Cao, and Diyi Yang. 2022.
\newblock \href {https://aclanthology.org/2022.lrec-1.626} {{SPORTSINTERVIEW}: A large-scale sports interview benchmark for entity-centric dialogues}.
\newblock In \emph{Proceedings of the Thirteenth Language Resources and Evaluation Conference}, pages 5821--5828, Marseille, France. European Language Resources Association.

\bibitem[{Truong and Lauw(2019)}]{truong-etal-2019-multimodal}
Quoc-Tuan Truong and Hady Lauw. 2019.
\newblock \href {https://doi.org/10.1145/3308558.3313463} {Multimodal review generation for recommender systems}.
\newblock In \emph{The World Wide Web Conference}, WWW '19, page 1864^^e2^^80^^931874, New York, NY, USA. Association for Computing Machinery.

\bibitem[{Vu et~al.(2020)Vu, Nguyen, Le, and Jiang}]{vu-etal-2020-multimodal}
Xuan-Son Vu, Thanh-Son Nguyen, Duc-Trong Le, and Lili Jiang. 2020.
\newblock \href {https://doi.org/10.18653/v1/2020.coling-main.37} {Multimodal review generation with privacy and fairness awareness}.
\newblock In \emph{Proceedings of the 28th International Conference on Computational Linguistics}, pages 414--425, Barcelona, Spain (Online). International Committee on Computational Linguistics.

\bibitem[{Wang et~al.(2023)Wang, Wei, Schuurmans, Le, Chi, Narang, Chowdhery, and Zhou}]{wang-etal-2022-selfconsistency}
Xuezhi Wang, Jason Wei, Dale Schuurmans, Quoc~V Le, Ed~H. Chi, Sharan Narang, Aakanksha Chowdhery, and Denny Zhou. 2023.
\newblock \href {https://openreview.net/forum?id=1PL1NIMMrw} {Self-consistency improves chain of thought reasoning in language models}.
\newblock In \emph{The Eleventh International Conference on Learning Representations}.

\bibitem[{Wei et~al.(2022)Wei, Wang, Schuurmans, Bosma, ichter, Xia, Chi, Le, and Zhou}]{wei-etal-2022-cot}
Jason Wei, Xuezhi Wang, Dale Schuurmans, Maarten Bosma, brian ichter, Fei Xia, Ed~Chi, Quoc~V Le, and Denny Zhou. 2022.
\newblock \href {https://proceedings.neurips.cc/paper_files/paper/2022/file/9d5609613524ecf4f15af0f7b31abca4-Paper-Conference.pdf} {Chain-of-thought prompting elicits reasoning in large language models}.
\newblock In \emph{Advances in Neural Information Processing Systems}, volume~35, pages 24824--24837. Curran Associates, Inc.

\bibitem[{Zhang et~al.(2021)Zhang, Zhang, Zaheer, and Ahmed}]{zhang-etal-2021-sutat}
Xinyuan Zhang, Ruiyi Zhang, Manzil Zaheer, and Amr Ahmed. 2021.
\newblock \href {https://doi.org/10.1609/aaai.v35i16.17703} {Unsupervised abstractive dialogue summarization for tete-a-tetes}.
\newblock In \emph{Proceedings of the AAAI Conference on Artificial Intelligence}, volume~35, pages 14489--14497.

\bibitem[{Zhao et~al.(2023)Zhao, Gella, Kim, Jin, Hazarika, Papangelis, Hedayatnia, Namazifar, Liu, and Hakkani-Tur}]{zhao-etal-2023-others}
Chao Zhao, Spandana Gella, Seokhwan Kim, Di~Jin, Devamanyu Hazarika, Alexandros Papangelis, Behnam Hedayatnia, Mahdi Namazifar, Yang Liu, and Dilek Hakkani-Tur. 2023.
\newblock \href {https://doi.org/10.18653/v1/2023.sigdial-1.28} {{``}what do others think?{''}: Task-oriented conversational modeling with subjective knowledge}.
\newblock In \emph{Proceedings of the 24th Annual Meeting of the Special Interest Group on Discourse and Dialogue}, pages 309--323, Prague, Czechia. Association for Computational Linguistics.

\bibitem[{Zhong et~al.(2022)Zhong, Liu, Xu, Zhu, and Zeng}]{zhong-etal-2022-dialoglm}
Ming Zhong, Yang Liu, Yichong Xu, Chenguang Zhu, and Michael Zeng. 2022.
\newblock Dialoglm: Pre-trained model for long dialogue understanding and summarization.
\newblock In \emph{Proceedings of the AAAI Conference on Artificial Intelligence}, volume~36, pages 11765--11773.

\bibitem[{Zhong et~al.(2021)Zhong, Yin, Yu, Zaidi, Mutuma, Jha, Awadallah, Celikyilmaz, Liu, Qiu, and Radev}]{zhong-etal-2021-qmsum}
Ming Zhong, Da~Yin, Tao Yu, Ahmad Zaidi, Mutethia Mutuma, Rahul Jha, Ahmed~Hassan Awadallah, Asli Celikyilmaz, Yang Liu, Xipeng Qiu, and Dragomir Radev. 2021.
\newblock \href {https://doi.org/10.18653/v1/2021.naacl-main.472} {{QMS}um: A new benchmark for query-based multi-domain meeting summarization}.
\newblock In \emph{Proceedings of the 2021 Conference of the North American Chapter of the Association for Computational Linguistics: Human Language Technologies}, pages 5905--5921, Online. Association for Computational Linguistics.

\bibitem[{Zhu et~al.(2021)Zhu, Liu, Mei, and Zeng}]{zhu-etal-2021-mediasum}
Chenguang Zhu, Yang Liu, Jie Mei, and Michael Zeng. 2021.
\newblock \href {https://doi.org/10.18653/v1/2021.naacl-main.474} {{M}edia{S}um: A large-scale media interview dataset for dialogue summarization}.
\newblock In \emph{Proceedings of the 2021 Conference of the North American Chapter of the Association for Computational Linguistics: Human Language Technologies}, pages 5927--5934, Online. Association for Computational Linguistics.

\bibitem[{Zou et~al.(2021)Zou, Zhu, Hu, Gui, and Zhang}]{zou-etal-2021-dams}
Yicheng Zou, Bolin Zhu, Xingwu Hu, Tao Gui, and Qi~Zhang. 2021.
\newblock \href {https://doi.org/10.18653/v1/2021.emnlp-main.7} {Low-resource dialogue summarization with domain-agnostic multi-source pretraining}.
\newblock In \emph{Proceedings of the 2021 Conference on Empirical Methods in Natural Language Processing}, pages 80--91, Online and Punta Cana, Dominican Republic. Association for Computational Linguistics.

\end{thebibliography}

\appendix
\section{Prompt Template}\label{sec:appendix-a}
\subsection{Prompt for Interview Dialogue}\label{sec:appendix-a-interview}
Table \ref{tb:prompt-interview} shows a prompt template for interview dialogues. [PRODUCT\_NAME] is a placeholder for the product title, which will be replaced with the product title selected by the participant. [MAX\_QUESTION] and [MIN\_QUESTION] are placeholders for the maximum and minimum number of dialogue turns. In our experiments, we used 15 and 8, respectively.

\subsection{Prompt for Review Generation}\label{sec:appendix-a-reviewgen}
Table \ref{tb:prompt-review} shows a prompt template for review generation. Similar to that for interviewing, [PRODUCT\_NAME] is a placeholder for the product title. [DIALOGUE] is a placeholder for the dialogue history, into which the interview dialogue history between our system and the participants is inserted.

\section{Baseline Details}\label{sec:appendix-b}
Table \ref{tb:questions-baseline} shows a prompt template for review generation. Similar to that for interviewing, [PRODUCT\_NAME] is a placeholder for the product title. [DIALOGUE] is a placeholder for the dialogue history, into which the interview dialogue history between our system and the participants is inserted.

\begin{table*}[t!]
  \centering
  \caption{Prompt template for interviewing.}
  \scalebox{0.98}{
    \setlength{\tabcolsep}{2mm}
    \renewcommand{\arraystretch}{1.0}
    \begin{tabularx}{\linewidth}{X}\hline
      Your role is ``interviewer'' and my role is ``interviewee''.\\
      About the product I am going to present, please elicit my impressions and opinions from me when I have touched it.\\
      \\
      Note the following statements.\\
      1. The interviewer elicits the interviewee's satisfaction and dissatisfaction (the positive and negative points) with the product in a well-balanced and detailed.\\
      2. In response to the interviewee's response, the interviewer asks more in-depth questions about the aspect or elicits feedback about other aspects of the product.\\
      3. Be sure to attach the name of your role at the beginning of your utterance. Since your role is ``interviewer'', your generation should begin with ``Interviewer:''.\\
      4. Don't generate interviewee's utterances.\\
      5. Add ``[Wait\_for\_Response]'' at the end of your utterance and wait for my response.\\
      6. You must ask at least [MIN\_QUESTION] questions. In other words, the dialogue must continue for [MIN\_QUESTION] or more turns.\\
      7. Having fulfilled the 6th statement, you can terminate the interview at your discretion. However, the interview must be completed within [MAX\_QUESTION] turns.\\
      8. When you terminate the intervew, add ``[End\_of\_Interview]'' at the end of your utterance.\\
      Now, please elicit my impressions and opinions about the following product from me.\\
      {[PRODUCT\_NAME]}\\ \hline
      \label{tb:prompt-interview}
    \end{tabularx}
  }
\end{table*}

\begin{table*}[t!]
  \centering
  \caption{Prompt template for review generation.}
  \scalebox{0.98}{
    \setlength{\tabcolsep}{2mm}
    \renewcommand{\arraystretch}{1.0}
    \begin{tabularx}{\linewidth}{X}\hline
      {[DIALOGUE]}\\
      \\
      The above is a dialogue about ``[PRODUCT\_NAME]'' between the interviewer and the interviewee who has touched on this product.\\
      \\
      Write a customer review about the product as if written by the interviewee, by briefly summarizing the important information mentioned in the above interview, such as the good and bad points of the product and the interviewee's experience with it.\\
      Do not output the review's title.\\
      The following is a body of the product review of the product written by the interviewee: \\ \hline
    \end{tabularx}
    \label{tb:prompt-review}
  }
\end{table*}

\begin{table*}[t!]
	\centering
	\caption{Questions asked by the baseline system}
  \scalebox{0.98}{
    \renewcommand{\arraystretch}{1.0}
    \begin{tabularx}{\linewidth}{lX}\hline
      Q-1 & First, could you tell me about the features and functions of this product? What kind of product is this?\\
      Q-2 & What made you decide to purchase this product?\\
      Q-3 & If you have any points that you like or are satisfied with this product, please tell me in detail.\\
      Q-4 & What are the advantages of this product compared to other products?\\
      Q-5 & If you have any dissatisfaction with this product or areas for improvement for this product, please tell me in detail.\\
      Q-6 & What are the disadvantages of this product compared to other products?\\
      Q-7 & Who would this product be suitable for?\\
      Q-8 & Is this product worth the price? Also, why do you think so?\\
      Q-9 & Finally, do you have any requests or impressions about the product?\\ \hline
    \end{tabularx}
  }
  \label{tb:questions-baseline}
\end{table*}

\end{document}